\documentclass{aastex}          
\usepackage{spr-astr-addons}    
\usepackage{epsfig}
\usepackage{graphicx}
\usepackage{color}

\begin{document}
%
\title{Observation of the period ratio $P_{1}/P_{2}$ of transversal oscillations in solar macro-spicules}

\shorttitle{Magnetohydrodynamic waves in solar spicules}
\shortauthors{Ebadi et al.}

\author{H.~Ebadi, and M.~Khoshrangbaf}
\affil{Astrophysics Department, Physics Faculty,
University of Tabriz, Tabriz, Iran\\
e-mail: \textcolor{blue}{hosseinebadi@tabrizu.ac.ir}}

\begin{abstract}
We analyze the time series of oxygen line profiles (O$\textsc{vi}$ 1031.93 ${\AA}$ and O$\textsc{vi}$ 1037.61 ${\AA}$)
obtained from SUMER/SOHO on the solar south limb. We calculated Doppler shifts and
consequently Doppler velocities in three heights $4^{''}$, $14^{''}$, and $24^{''}$ from the limb on a coronal hole region.
Then, we performed wavelet analysis with Morlet wavelet transform to determine the periods of fundamental mode and
its first harmonic mode. The calculated period ratios have departures from its canonical value of $2$.
The density stratification and magnetic twist are two main factors which may cause these departures.

\end{abstract}


\section{Introduction}
\label{sec:intro}
Observation of oscillations in solar spicules may be used as an indirect evidence of energy transport from the photosphere towards the corona.  Transverse motion of spicule axis can be observed by both, spectroscopic and imaging observations.  The periodic Doppler shift of spectral lines have been observed from ground based coronagraphs \citep{nik67,Kukh2006,Tem2007}.  But Doppler shift oscillations with period of $\sim\!\!5$ min also have been observed on the SOlar and Heliospheric Observatory (\emph{SOHO}) by \citet{xia2005}.  Direct periodic displacement of spicule axes have been found by imaging observations on Solar Optical Telescope (SOT) on \emph{Hinode\/} \citep{De2007,Kim2008,he2009}.

The observed transverse oscillations of spicule axes were interpreted by kink
\citep{nik67,Kukh2006,Tem2007,Kim2008,Ebadi2012a} and Alfv\'{e}n \citep{De2007} waves.
All spicule oscillations events are summarized in a recent review by \citet{Tem2009}.
They suggested that the observed oscillation periods can be formally divided in two groups:
those with shorter periods ($<\!\!2$ min) and those with longer periods ($\geqslant \!\!2$ min) \citep{Tem2009}.
The most frequently observed oscillations lie in the period ranges of $3$--$7$ min and $50$--$110$ s.
Additionally, very long spicules, called as macrospicules by \citet{Bochlin1975}
with a typical length of up to 40 Mm are frequently observed mostly near the
polar regions as reported by, e.g., \citet{Pike1997,Doyle2005,Scullion2009,Murawski2011}.

One of the most important functions of coronal seismology is determining the period ratio $P_{1}/P_{2}$ between the period
$P_{1}$ of the fundamental mode and the period $P_{2}$ of its first harmonic \citep{Andries2009,Karami2012,Orza2012,Erdelyi2013}.
Different factors such as the effect of density stratification \citep{Andries2009} and magnetic twist \citep{Karami2009}
can cause the deviation of the period ratio from its canonical value of $2$. The observed values of this ratio
in coronal loops is either smaller or larger than $2$ \citep{Verwichte2004,Andries2009}.
\citet{Srivastava2008} Using simultaneous high spatial and temporal resolution H$\alpha$
observations studied the oscillations in the relative intensity to explore the
possibility of sausage oscillations in the chromospheric cool post-flare loop.
They used the standard wavelet tool, and find $P_{1}/P_{2} \sim 1.68$. They suggested
that the oscillations represent the fundamental and the first harmonics
of the fast-sausage waves in the cool post-flare loop.
\citet{Verwichte2004} have detected interesting phenomenon
of simultaneous existence of fundamental and first harmonics
of fast-kink oscillations (see also \citet{De Moortel2007,Van Doorsselaere2007,Tem2013a}).
However, the ratio between the periods of fundamental and first harmonics
$P_{1}/P_{2}$ was significantly shifted from $2$, which later was explained
as a result of longitudinal density stratification in the loop. The rate of the
shift allows us to estimate the density scale height in coronal loops,
which can be a few times larger compared to its hydrostatic value.

Observed oscillation periods can be used to estimate the Alfv\'en speed and consequently magnetic
field strength in macro-spicules \citep{Tem2013b}.

The mentioned studies in the previous paragraph are all devoted to the coronal loop transversal oscillations.
To my knowledge there is no any work related to the period ratio of spicules oscillations. So, the present study
is an attempt to check this ratio observationally. We will study the same problem theoretically in our future works.

\section{Observations and data processing}
\label{sec:observations}
SUMER is a high-resolution normal incidence spectrograph operating
 in the range 666-1610\textbf{$\sim$}{\AA}  (first order) and
333-805\textbf{$\sim$}{\AA} (second order). The
angular pixel size is $\sim$\H{1}. The spectral
pixel size depends slightly on the wavelength.
Contriving normally allows sub-pixel resolution. It can
vary from about 45 m{\AA}/pixel at 800\textbf{$\sim$}{\AA} to
about 41\textbf{$\sim$}m{\AA}/pixel at 1600\textbf{$\sim$}{\AA} \citep{Wilhelm1995}.

\begin{figure}
\centering
\includegraphics[width=8cm]{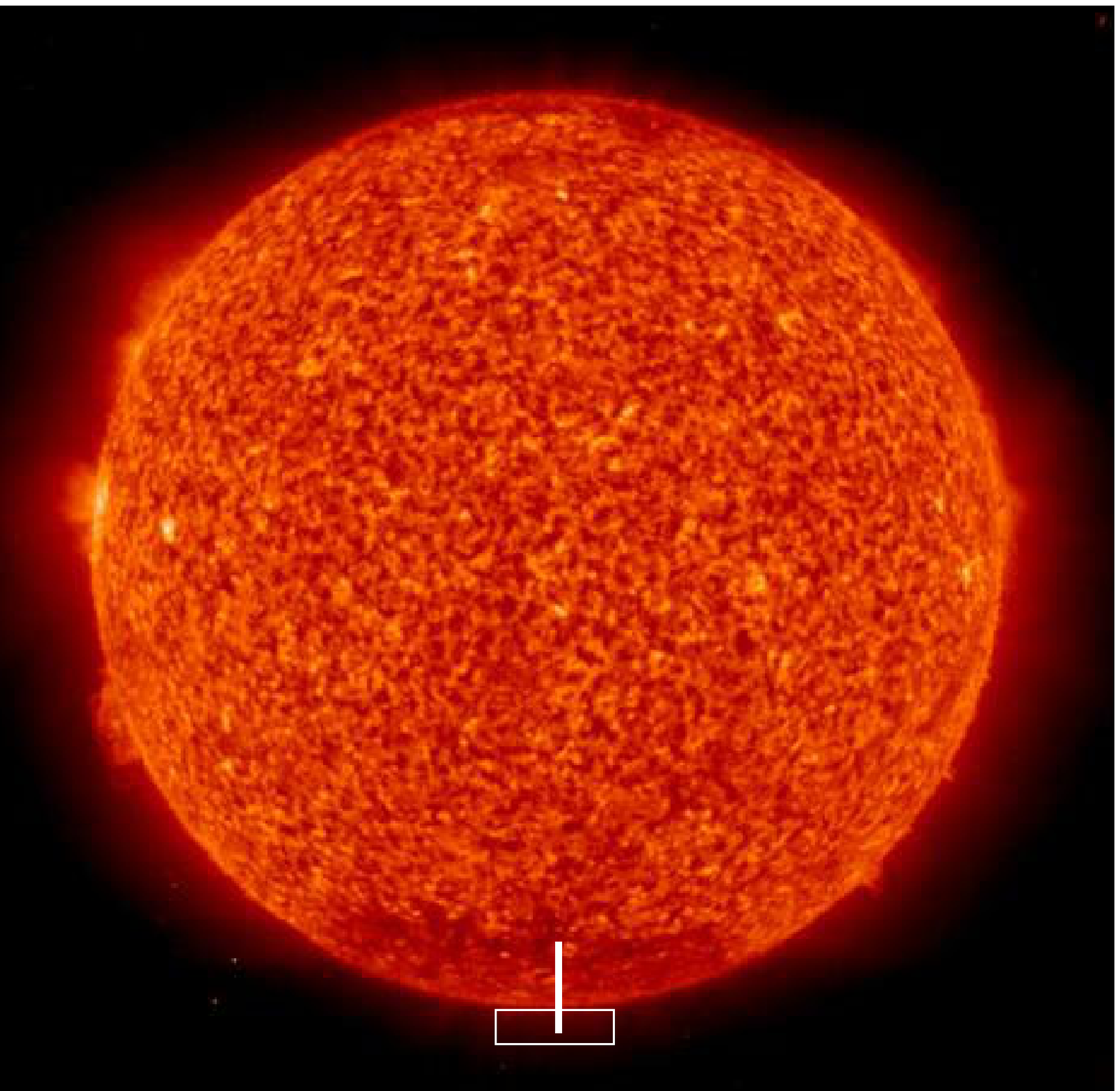}
\includegraphics[height=2.5cm, width=8cm]{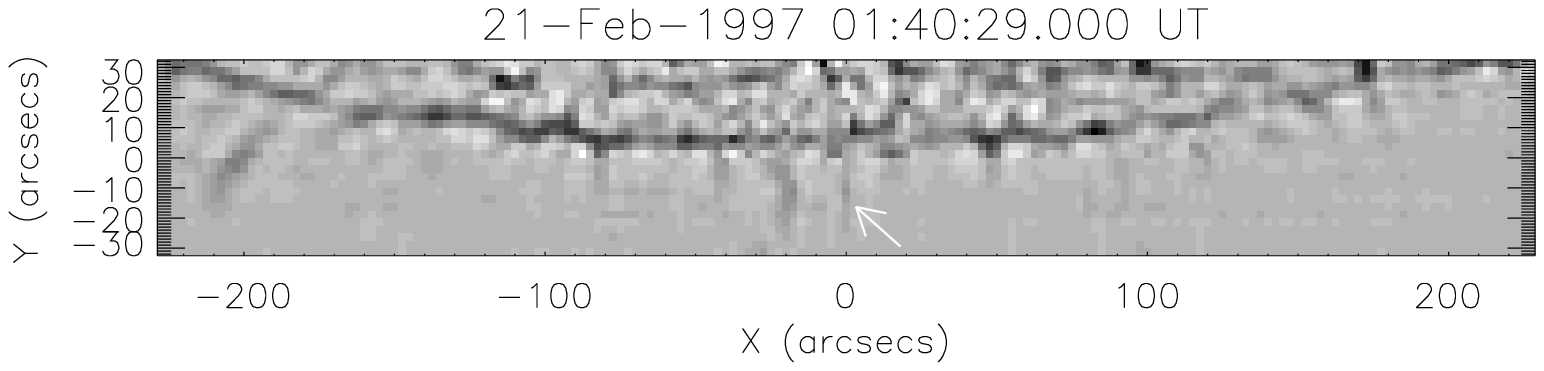}
\caption{Image of the studied region which were observed by 304 {\AA}
SOHO/EIT (top) on 21 February 1997. The slit is in the North\,--\,South
direction. The rectangular shows the region studied. We used the ``madmax'' algorithm
to enhance the finest structures (down). The white arrow shows the studied macro-spicule. \label{fig1}}
\end{figure}

A coronal hole region in the south pole of the sun was
observed with SUMER (detector B) on 21 Feb 1997. The pointing
coordinates were X = 0 $^{''}$, Y = -985$^{''}$. The slit,
which was used for observations, has the dimensions of
 0.3$^{''}$$\times$120$^{''}$. The observation was performed
from 01:36 UT to 01:52 UT and the exposure time was 15 seconds.
In Figure~\ref{fig1} we presented the images of the studied region which
were observed by 304 {\AA} SOHO/EIT (top) on 21 February 1997. The rectangular
shows the region of south limb macro-spicules. We used the ``madmax'' algorithm
to enhance the finest structures \citep{Koutchmy89}. As it is clear from down panel of Figure~\ref{fig1},
the length of the studied spicule is $25$ Mm which means that the studied spicule is macro-spicule.
Since the EIT images have a fixed pixel resolution of $2.5$ arc\,sec so,
the down panel of Figure~\ref{fig1} is the best quality after image processing techniques.

The raw data have been initially processed applying the standard
procedures for flatfield, deadtime and destretching
correction which can be found in the Solar Software (SSW)
database. We performed the radiometric calibration, so the
specific intensity unit is W m$^{-2}$ sr$^{-1}$ \AA$^{-1}$ through
this analysis\citep{Ebadi2007,Ebadi2009}.

\begin{figure}[!h]
\epsscale{1.0}
\plotone{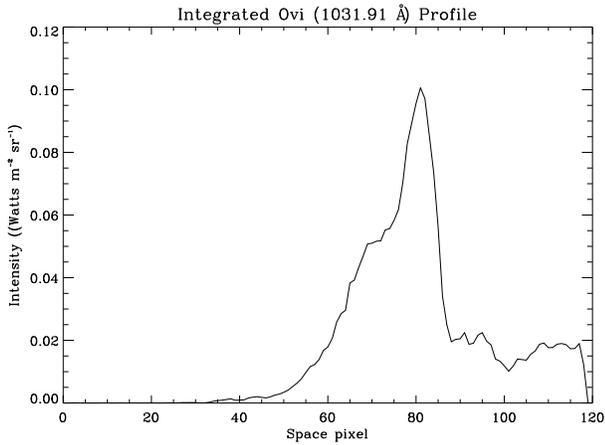}
\caption{The integrated intensity along the SUMER slit for O$\textsc{vi}$ (1031.93 ${\AA}$) line.
The spicule region is lied from pixel $87$ to $117$.\label{fig2}}
\end{figure}

\begin{figure}[!h]
\epsscale{1.0}
\plotone{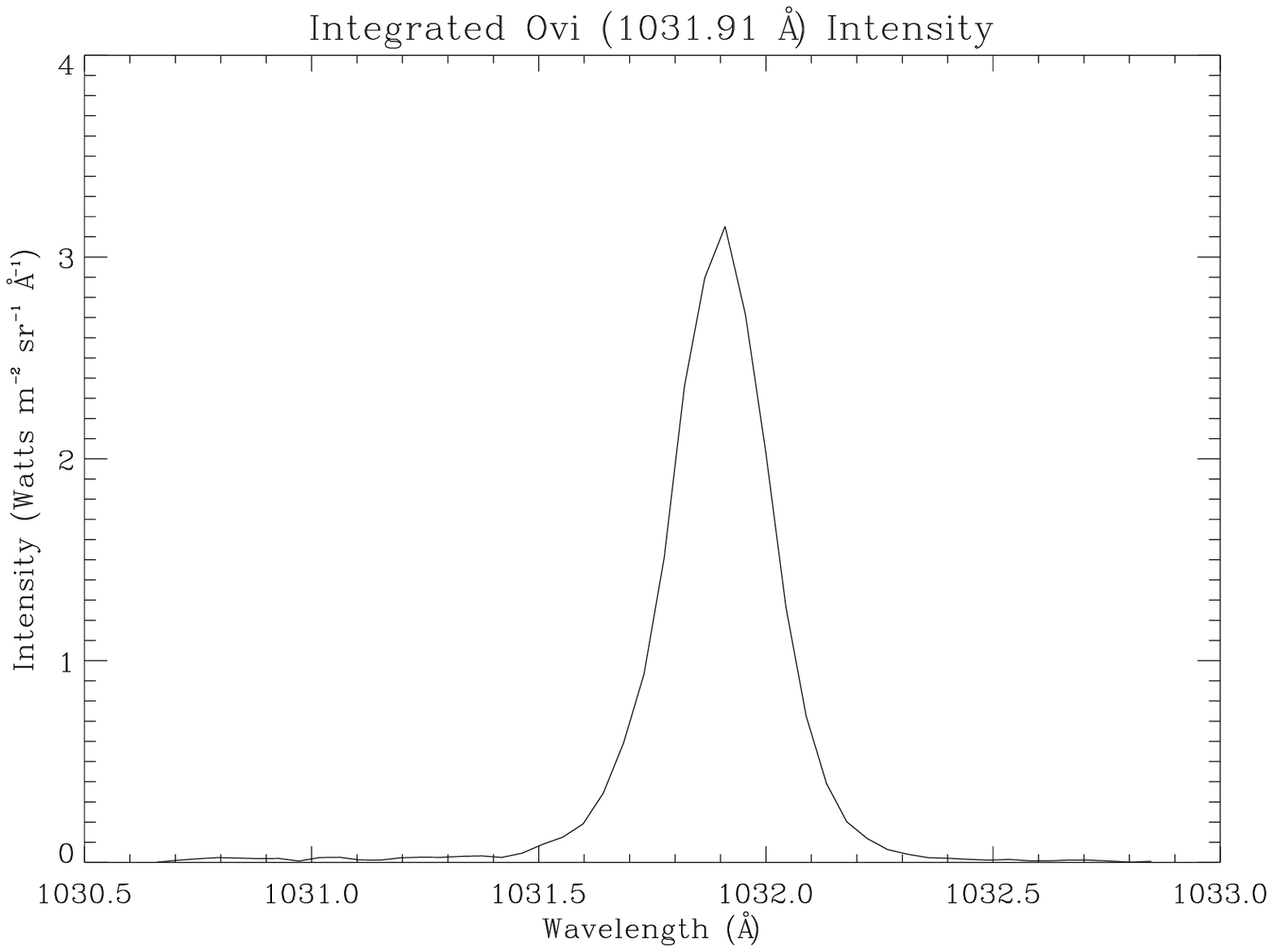}
\caption{The integrated profile of O$\textsc{vi}$ (1031.93 ${\AA}$) line.\label{fig3}}
\end{figure}

We calculated the integrated intensity for O$\textsc{vi}$ (1031.93 ${\AA}$) line along the SUMER slit.
The limb is located in pixel number $86$ and the spicule region is lied from pixel $87$ to $117$ which is shown
in Figure~\ref{fig2}. Moreover, we plotted integrated profile  of O$\textsc{vi}$ (1031.93 ${\AA}$) line in Figure~\ref{fig3}.


\section{Results and Discussion}
\label{sec:results}
\begin{figure*}[!h]
\epsscale{1.5}
\plotone{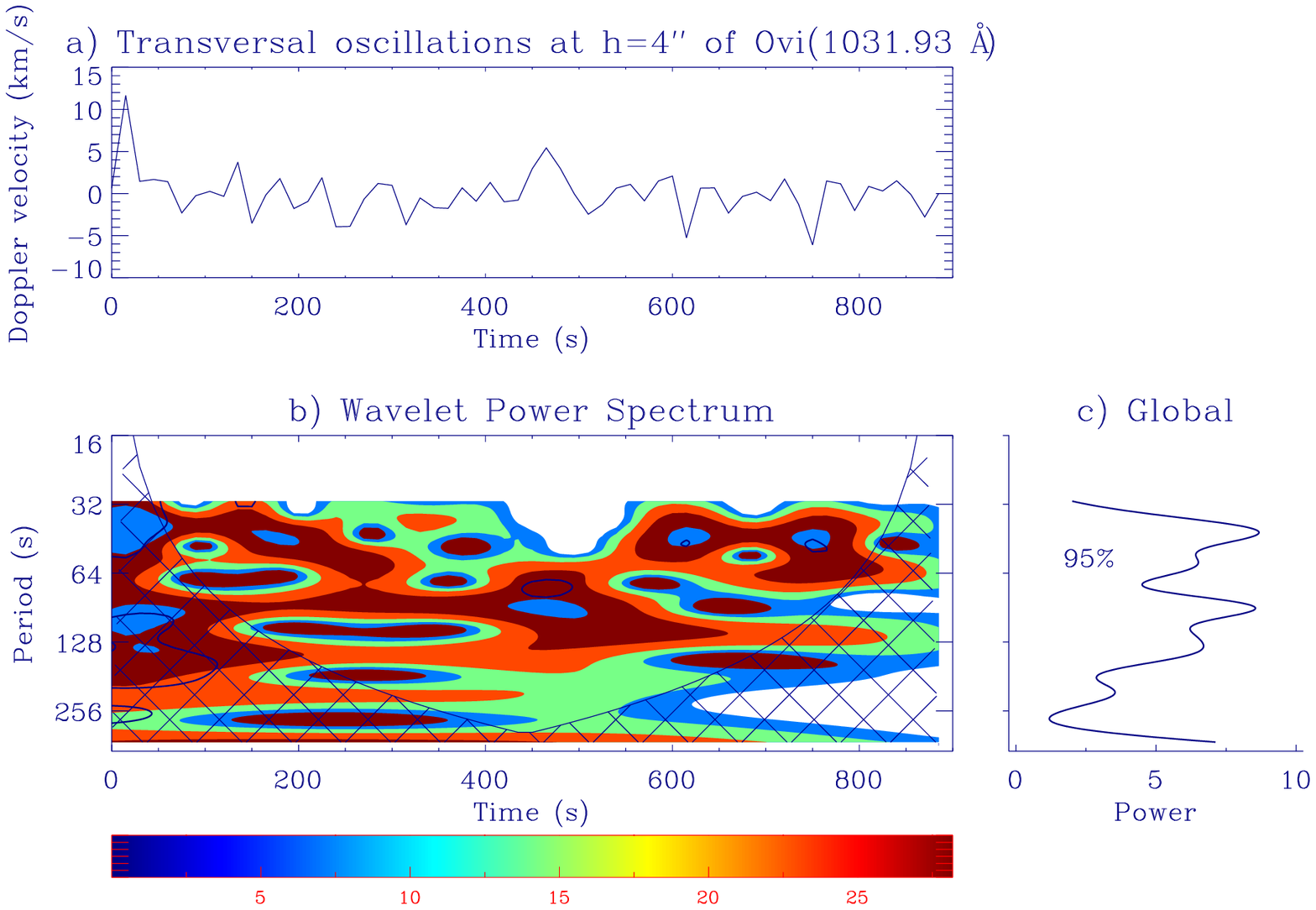}
\caption{a. Doppler velocity variations of the studied spicule $4^{''}$
above the limb in O$\textsc{vi}$ (1031.93 ${\AA}$) line.
b. The wavelet power spectrum. The contour levels are chosen so that $75\%$, $50\%$, $25\%$,
and $5\%$ of the wavelet power is above each level, respectively.
The cross-hatched region is the cone of influence, where zero padding has reduced the variance.
c. The global wavelet power spectrum.\label{fig4}}
\end{figure*}

\begin{figure*}[!h]
\epsscale{1.5}
\plotone{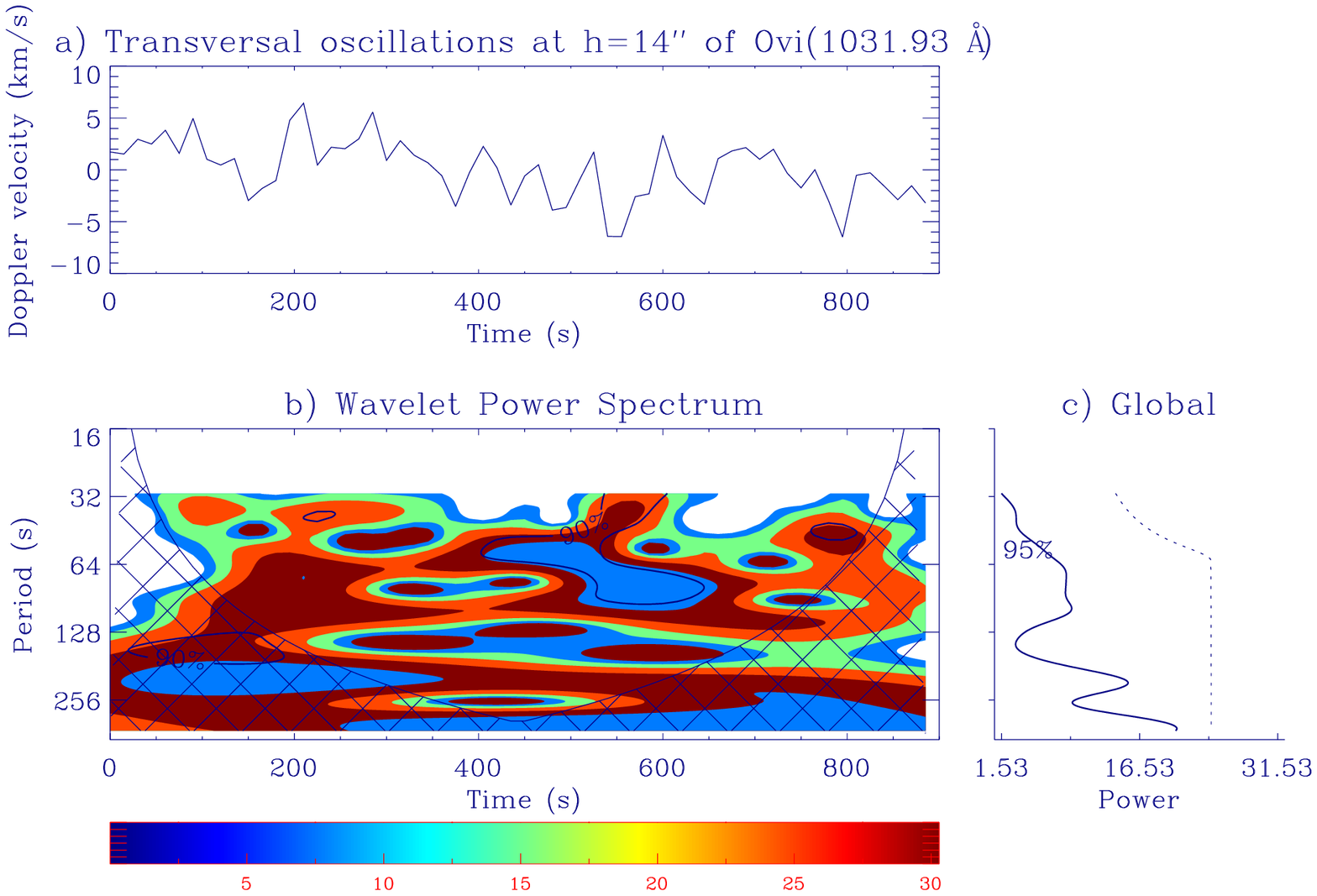}
\caption{The same as in Fig.~\ref{fig3} but $14^{''}$ above the limb.\label{fig5}}
\end{figure*}

\begin{figure*}[!h]
\epsscale{1.5}
\plotone{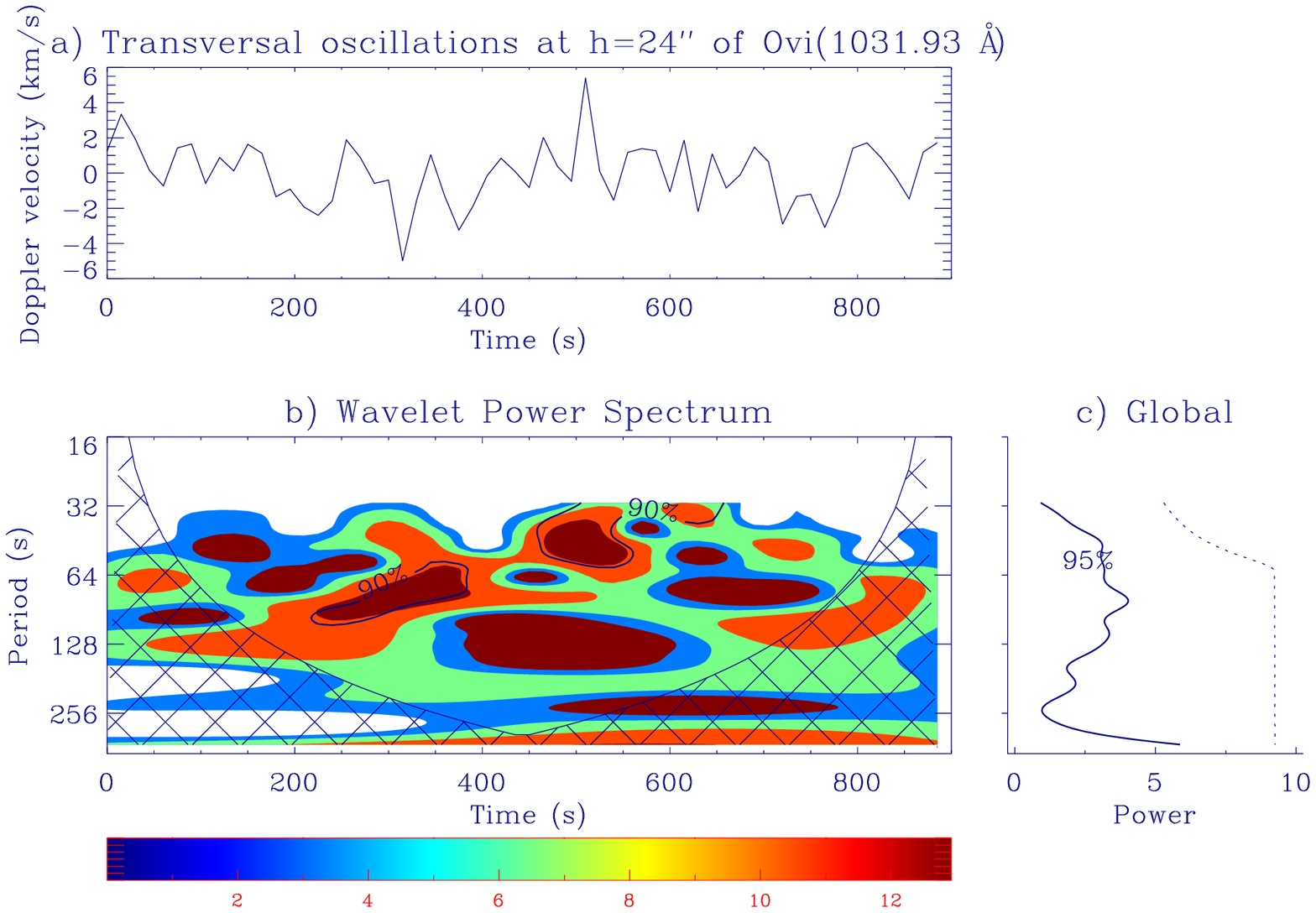}
\caption{The same as in Fig.~\ref{fig3} but $24^{''}$ above the limb.\label{fig6}}
\end{figure*}

\begin{figure*}[!h]
\epsscale{1.5} \plotone{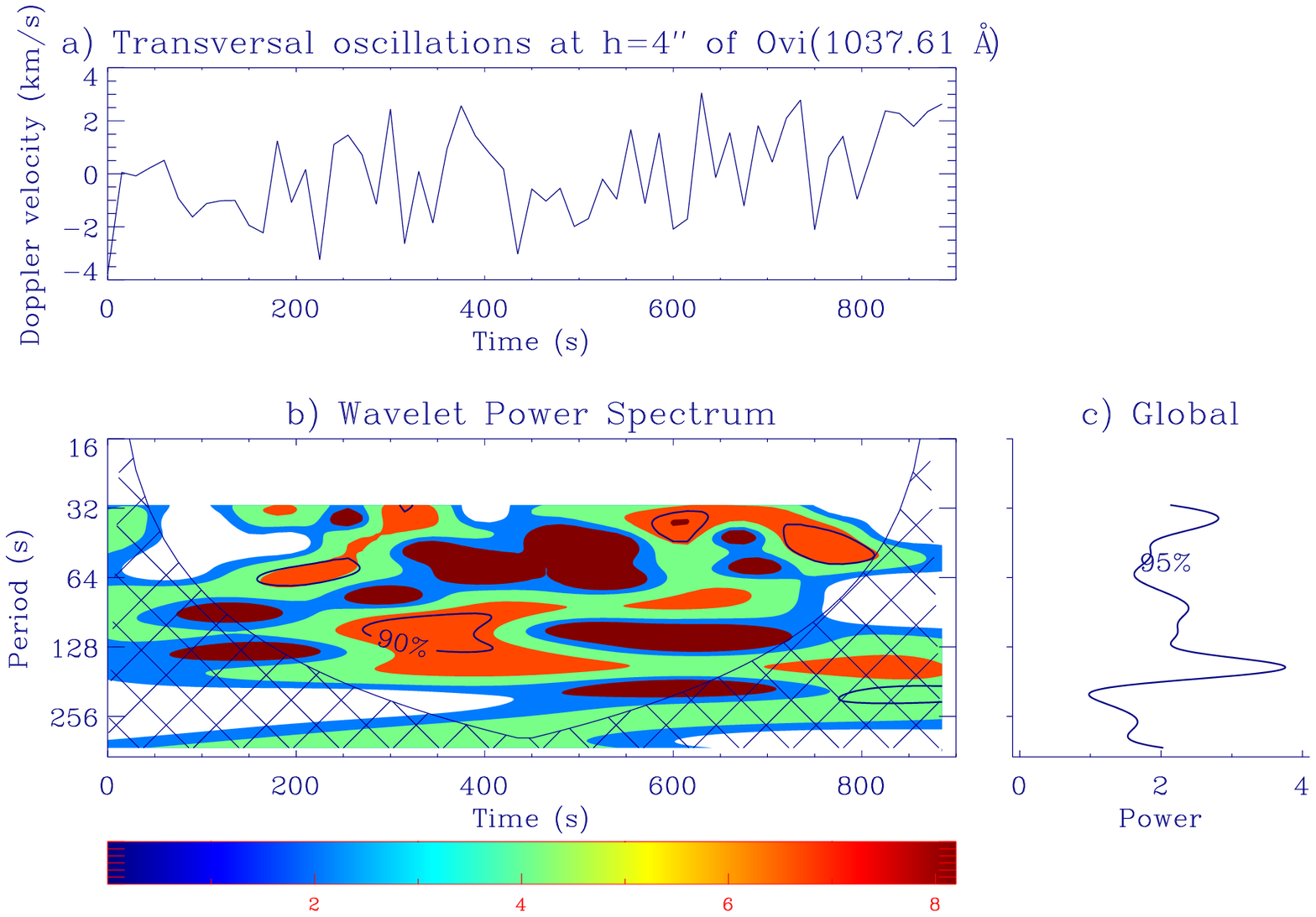} \caption{a. Doppler velocity variations of the studied spicule $4^{''}$
above the limb in O$\textsc{vi}$ (1037.61 ${\AA}$) line.
b. The wavelet power spectrum. The contour levels are chosen so that $75\%$, $50\%$, $25\%$,
and $5\%$ of the wavelet power is above each level, respectively.
The cross-hatched region is the cone of influence, where zero padding has reduced the variance.
c. The global wavelet power spectrum.
\label{fig7}}
\end{figure*}

\begin{figure*}[!h]
\epsscale{1.5} \plotone{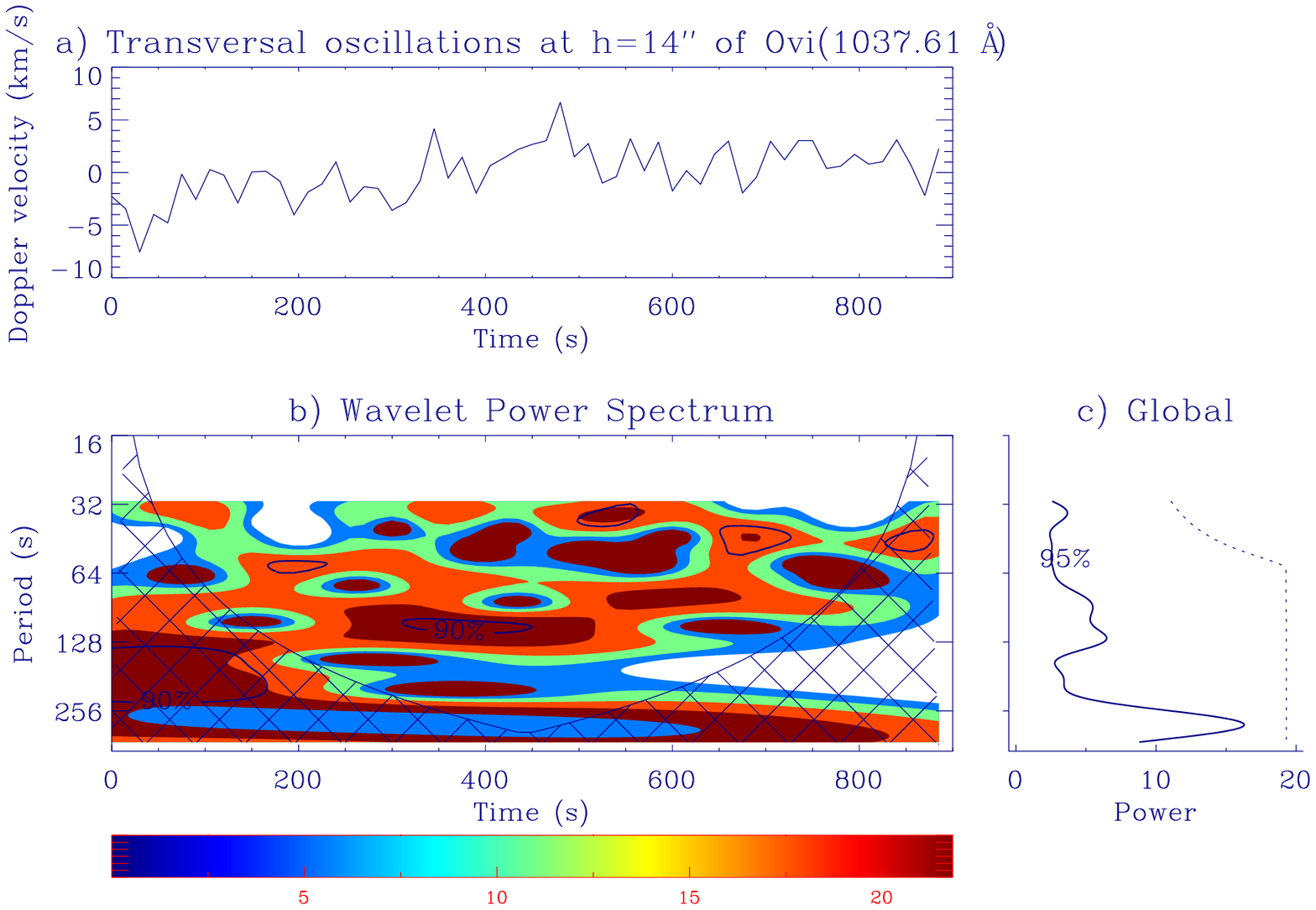} \caption{The same as in Fig.~\ref{fig6} but $14^{''}$ above the limb.\label{fig8}}
\end{figure*}

\begin{figure*}[!h]
\epsscale{1.5} \plotone{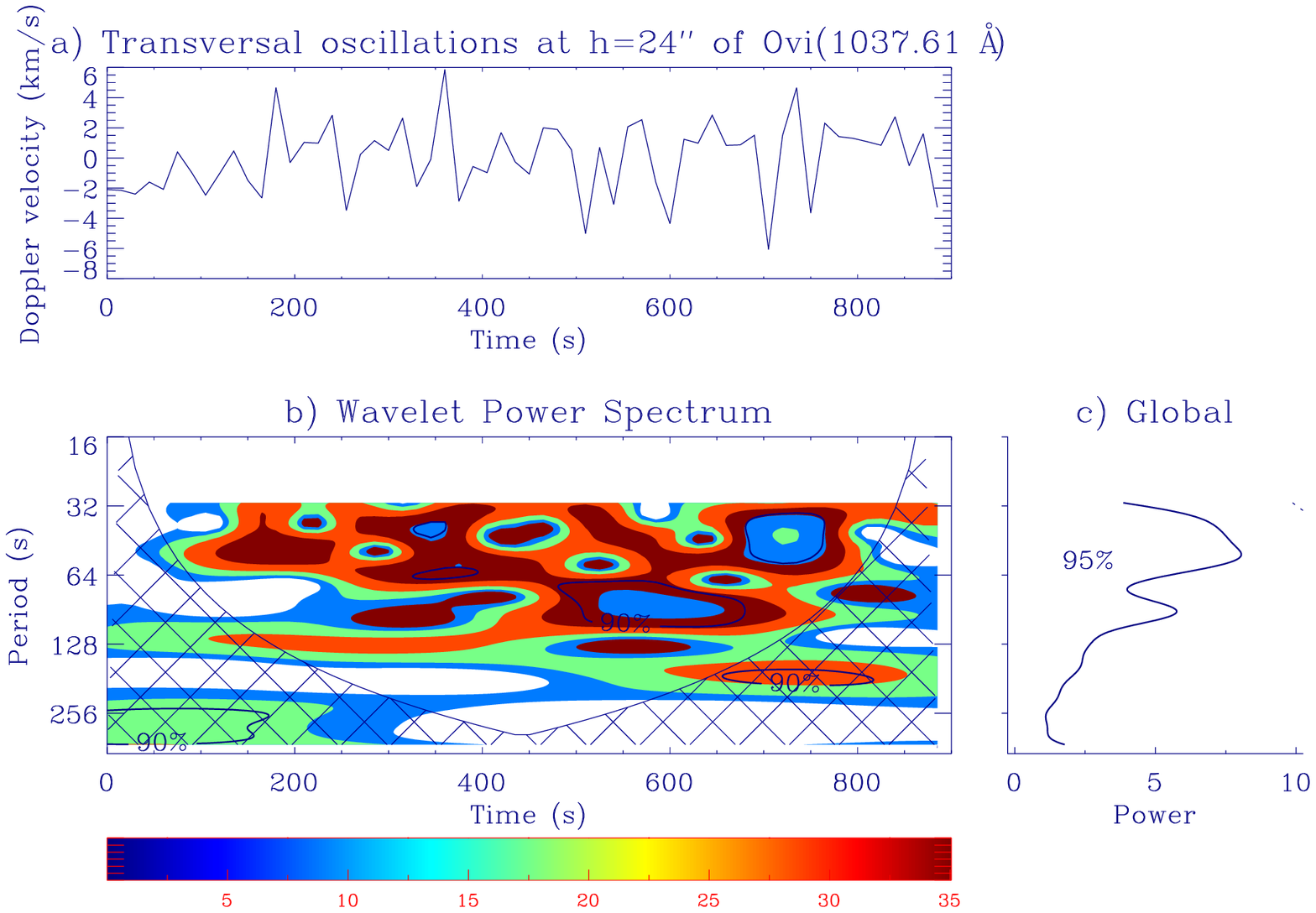} \caption{The same as in Fig.~\ref{fig6} but $24^{''}$ above the limb.\label{fig9}}
\end{figure*}

We analyze O$\textsc{vi}$ (1031.93 ${\AA}$) and O$\textsc{vi}$ (1037.61 ${\AA}$) line profiles
from the time series by fitting to a Gaussian. Then we calculated Doppler shifts and
consequently Doppler velocities \citep{Tem2007}. we used the two stable photospheric neutral oxygen
emission lines (i.e. O$\textsc{i}$ (1027.43 ${\AA}$) and O$\textsc{i}$ (1028.16 ${\AA}$)) that happen to be in the same
spectral window with the O$\textsc{vi}$ lines. Doppler velocities and proper wavelet analysis results
are presented in Figures~\ref{fig4},~\ref{fig5}, and~\ref{fig6} for O$\textsc{vi}$ (1031.93 ${\AA}$). We perform wavelet
analysis with Morlet wavelet transform in three heights for both lines. On the other hand, wavelet analysis
results for the line O$\textsc{vi}$ (1037.61 ${\AA}$) are showed in Figures~\ref{fig7},~\ref{fig8}, and~\ref{fig9}.
The wavelet power spectrum, the cone of influence, and the global wavelet power spectrum are plotted in each figure.
The contour levels are chosen so that $75\%$, $50\%$, $25\%$,
and $5\%$ of the wavelet power is above each level, respectively.
We determined the fundamental mode and its first harmonic periods for each line and in three heights ($4^{''}$, $14^{''}$, and $24^{''}$ from the limb).
For comparison we presented results in Table 1 for both lines.

\begin{table}[htbp]
\caption{$P_{1}$(fundamental mode period), $P_{2}$(first harmonic period), and $P_{1}/P_{2}$
(fundamental to its first harmonic period ratio) are presented for both oxygen lines.}
   \label{tabfreqoliver}
\vspace{0.6cm}
\begin{tabular}{c c c c c c c c}
   \hline
   line                         &height               & $P_{1}$(s) & $P_{2}$(s) & $P_{1}/P_{2}$ \\
   \hline
   O$\textsc{vi}(1031.93{\AA}$) & $4^{''}$                  &    96      &    45      &      2.13     \\
   \hline
   O$\textsc{vi}(1031.93{\AA}$) & $14^{''}$                 &    100     &    48      &      2.08     \\
   \hline
   O$\textsc{vi}(1031.93{\AA}$) & $24^{''}$                 &    110     &    50      &      2.2      \\
   \hline
   O$\textsc{vi}(1037.61{\AA}$) & $4^{''}$                  &    108     &    52      &      2.07       \\
   \hline
   O$\textsc{vi}(1037.61{\AA}$) & $14^{''}$                 &    118     &    58      &      2.03       \\
   \hline
   O$\textsc{vi}(1037.61{\AA}$) & $24^{''}$                 &    100     &    49      &      2.04       \\
   \hline
\end{tabular}
\end{table}

As it is clear from the Table 1, the fundamental mode and its first harmonic period ratios have departures from
its canonical value of $2$. They are greater than $2$ in both lines in $4^{''}$, $14^{''}$, and $24^{''}$ from the limb.
The fundamental mode periods are determined as $96-118$ s and its first harmonic periods are lied in the range $45-58$ s.
These periods are in good agreement with the results of previous works in this area \citep{Kukh2006,Ebadi2012a,Ebadi2012b,Ebadi2013}.
It should be noted that these periods and their ratios are related to a unique macro-spicule. It is
shown in Figure~\ref{fig1} that the SUMER slit covers only one macro-spicule.

The density stratification and magnetic twist are two main factors which make the period ratio departures
from its canonical value of $2$. These two factors are studied in spicules both observationally and
theoretically. The first estimation of spicule magnetic field by spicule seismology was done
by \citet{Tem2007}. \citet{Verth2011} based on SOT/\emph{Hinode} observations determined the variation of
magnetic field strength and plasma density along a spicule by seismology. They studied a kink wave
propagating along a spicule by estimating the spatial change in phase speed and velocity amplitude
as a novel approach. \citet{Suematsu2008,Tavabi2011} by using the SOT/\emph{Hinode} observations
reported twisted motions in spicules.

Since we determined the oscillation periods of fundamental mode and its first harmonic in macro-spicules,
it is possible to estimate the Alfv\'{e}n speed and consequently magnetic field strength through them.
Kink waves are transverse oscillations of magnetic tubes and the phase speed for a straight homogenous
tube can be written as:

\begin{equation}
\label{eq:phasespeed}
c_{k} = \frac{\lambda}{T_{obs}} = V_{A0}\sqrt{\frac{2}{1+\rho_{e}/\rho_{0}}},
\end{equation}
where $V_{A0}\equiv B_{0}/\sqrt{\mu_{0}\rho_{0}}$ is the Alfv\'{e}n speed inside the tube, $\lambda$ is the wavelength,
and $T_{obs}$ is the observed oscillation period.
$\rho_{e}$ and $\rho_{0}$ are the plasma density outside and inside of tube, respectively. 
The density in spicules is $3 \times 10^{-10}$ kg m$^{-3}$ and $\rho_{e}/\rho_{0}\simeq 0.02$ \citep{Tem2009,Ebadi2012a}. 
For the fundamental mode, $\lambda=2L$ ($L$ is the length of the macro-spicule which determined as $25$ Mm in this work). 
The mean fundamental mode period is determined as $\sim 105$ s. By using these parameters in Equation~\ref{eq:phasespeed} 
the Alfv\'{e}n speed and magnetic field strength is estimated as $\sim340$ km/s and $\sim65$ G in macro-spicules, respectively. 

\section{conclusion}
\label{sec:concl}
We analyze the time series of O$\textsc{vi}$ (1031.93 ${\AA}$) and O$\textsc{vi}$ (1037.61 ${\AA}$)
line profiles obtained from SUMER/SOHO in order to uncover the oscillations in the solar macro-spicules.
We concentrate on particular coronal hole region which contains the macro-spicules and found that
their axis undergo quasi-periodic transverse displacement. This is done by calculating Doppler shifts and
consequently Doppler velocities in three heights $4^{''}$, $14^{''}$, and $24^{''}$ from the limb.
By performing wavelet analysis with Morlet wavelet transform in three heights we determine the
fundamental mode and its first harmonic periods and their ratios. Our findings show small departures of
this value from its canonical value of $2$ in both lines and three mentioned heights. In other words,
they are greater than $2$ in both lines in $4^{''}$, $14^{''}$, and $24^{''}$ from the limb.
These departures may be caused by the density stratification and magnetic twist which is observed in spicules. 
Observed oscillation periods are used to estimate the Alfv\'en speed and consequently magnetic
field strength in macro-spicules as $\sim340$ km/s and $\sim65$ G, respectively.

\acknowledgments
The authors thank anonymous referee for his/her useful and instructive comments in improving the document. 
The authors are grateful to the \mbox{\emph{SOHO}} Team for providing the observational data. SUMER is financially supported by
DLR, CNES, NASA and the ESA PRODEX program (Swiss contribution).
SOHO is a mission of international cooperation between ESA and
NASA. We appreciate the ISSI support in the frame of the
"Spectroscopy and Imaging of coronal hole spicules from Space" Team.

\makeatletter
\let\clear@thebibliography@page=\relax
\makeatother

\end{document}